\documentclass[aps,prb]{revtex4-2}

\usepackage{graphicx}
\usepackage{upgreek}
\usepackage{amsmath}
\usepackage{amssymb}
\usepackage{dcolumn}
\usepackage{chngpage}

\usepackage{CJK}

\usepackage{graphicx}
\usepackage{dcolumn}
\usepackage{bm}
\usepackage{amssymb} 
\usepackage{subfigure}
\usepackage{color}
\usepackage{epstopdf}
\usepackage{tabularx}

\begin{document}

\title{Emergence of spin singlets with inhomogeneous gaps in the kagome Heisenberg antiferromagnets Zn-barlowite and herbertsmithite}

\author{Jiaming Wang$^{1}$}
\author{Weishi Yuan$^{1}$}
\author{Philip M. Singer$^{2}$}
\author{Rebecca W. Smaha$^{3,4}$}
\author{Wei He$^{3,5}$}
\author{Jiajia Wen$^{3}$}
\author{Young S. Lee$^{3,6}$}
\author{Takashi Imai$^{1,7}$}
\email[Corresponding author: ]{imai@mcmaster.ca}

\affiliation{$^{1}$Department of Physics and Astronomy, McMaster University, Hamilton, Ontario, L8S 4M1, Canada}
\affiliation{$^{2}$Department of Chemical and Biomolecular Engineering, Rice University, 6100 Main St., Houston, TX 77005, USA}
\affiliation{$^{3}$Stanford Institute for Materials and Energy Sciences, SLAC National Accelerator Laboratory, Menlo Park, CA 94025, USA}
\affiliation{$^{4}$Department of Chemistry, Stanford University, Stanford, CA 94305, USA}
\affiliation{$^{5}$Department of Materials Science and Engineering, Stanford University,Stanford, CA 94305, USA}
\affiliation{$^{6}$Department of Applied Physics, Stanford University, Stanford, CA 94305, USA}
\affiliation{$^{7}$Brockhouse Institute for Materials Research, McMaster University, Hamilton, Ontario, L8S 4M1, Canada}






%


\maketitle


{\bf The kagome Heisenberg antiferromagnet formed by frustrated spins arranged in a lattice of corner-sharing triangles  is a prime candidate for hosting a quantum spin liquid (QSL) ground state consisting of entangled spin singlets~\cite{Balents2010}.  But the existence of various competing states makes a convincing theoretical prediction of the QSL ground state difficult~\cite{Broholm2020}, calling for experimental clues from model materials.  The kagome lattice materials Zn-barlowite ZnCu$_{3}$(OD)$_{6}$FBr~\cite{Feng2017,Smaha2020,Tustain2020} and herbertsmithite ZnCu$_{3}$(OD)$_{6}$Cl$_2$~\cite{Shores2005,Helton2007,Han2012,Fu2015,Khunita2020} do not exhibit long range order, and they are considered the best realizations of the kagome Heisenberg antiferromagnet known to date.  Here we use $^{63}$Cu nuclear quadrupole resonance combined with the inverse Laplace transform (ILT)~\cite{Song2002,Singer2020,Papawassiliou2020}  to probe locally the inhomogeneity of delicate quantum ground states affected by disorder~\cite{Singh2010,Shimokawa2015,Kimchi2018,Kawamura2019}.  We present direct evidence for the gradual emergence of spin singlets with spatially varying excitation gaps, but even at temperatures far below the super-exchange energy scale their fraction is limited to approximately 60\% of the total spins.  Theoretical models~\cite{Yan2011,Liao2017}  need to incorporate the role of disorder to account for the observed inhomogeneously gapped behaviour.
}

Experimental investigations of the kagome Heisenberg antiferromagnet often encounter complications due to undesirable long range order, spin freezing, or deviations from the ideal kagome structure~\cite{Hiroi2001,Matan2010,Ranjith2018,Klanjsek2017}.  In contrast, Zn-barlowite ZnCu$_{3}$(OD)$_{6}$FBr~\cite{Feng2017} and herbertsmithite ZnCu$_{3}$(OD)$_{6}$Cl$_2$~\cite{Shores2005} consist of spin-1/2 moments of Cu$^{2+}$ ions arranged with the perfect kagome symmetry, and remain paramagnetic even at $T\sim10^{-4}J$, where $J\sim 190$~K is the Cu-Cu super-exchange interaction.  Accordingly, experiments on these materials are expected to provide a platform for testing the theoretical ideas developed for the  kagome Heisenberg antiferromagnet, such as determining whether the low energy spin excitations are gapped or gapless.

However, the direct comparison of theory to observations on these materials is complicated due to disorder.  Site-selective x-ray scattering experiments  demonstrates that their actual composition is (Zn$_{0.95}$Cu$_{0.05}$)Cu$_{3}$(OD)$_{6}$FBr~\cite{Smaha2020_PRM} and (Zn$_{0.85}$Cu$_{0.15}$)Cu$_{3}$(OD)$_{6}$Cl$_2$~\cite{Freedman2010}, because 5\% or 15\% excess Cu$^{2+}$ impurities occupy the non-magnetic Zn$^{2+}$ sites within the interlayers outside the kagome planes.  On the other hand, the upper bound for non-magnetic Zn$^{2+}$ defects occupying the Cu$^{2+}$ sites within the kagome planes is $\sim1$~\%~\cite{Freedman2010,Smaha2020_PRM}.  These interlayer defect concentrations are reproducible, and confirmed for samples within the same synthesis batches used in this study.  The interlayer defect Cu$^{2+}$ spins interact with their six nearest-neighbor Cu$^{2+}$ sites in the adjacent kagome planes,  and disrupt the spin liquid ground state in their vicinity.  In fact, earlier NMR Knight shift measurements at $^{2}$D~\cite{Imai2011} and $^{17}$O sites~\cite{Fu2015,Khunita2020} in herbertsmithite showed that the local spin susceptibility $\chi_\text{spin}$ is strongly enhanced in the immediate vicinity of the 15~\% interlayer Cu defects.  Inelastic neutron scattering also reveals a peculiar $\omega/T$-scaling behavior in spin fluctuations at low energies below $\omega\sim0.1J$, suggesting the critical roles played by disorder~\cite{Helton2010}.  In view of the general tendency in disordered antiferromagnets that spin singlets gradually emerge with spatially varying gaps~\cite{Bhatt-Lee1982}, disorder and the resultant inhomogeneity in these Kagome materials must be taken into account when interpreting the collective spin dynamics~\cite{Singh2010,Shimokawa2015,Kimchi2018,Kawamura2019}.

In this Letter, we introduce a new experimental approach to shed light on the magnetic inhomogeneity of these kagome materials.  We experimentally deduce the histogram $P(1/T_{1}^\text{Cu})$ of the distributed $^{63}$Cu nuclear spin-lattice relaxation rate $1/T_{1}^\text{Cu}$ from the nuclear spin recovery curve $M(t)$ based on the inverse Laplace transform (ILT)  $T_{1}$ analysis technique~\cite{Song2002,Singer2020,Papawassiliou2020}.  We uncover the gradual emergence below $\sim30$~K of the magnetically inert spin singlet Cu sites with spatially varying gaps, as evidenced by the gradual increase of the population of Cu sites with vanishing $1/T_{1}^\text{Cu}$.  Our observation of the robust spin singlet signature but only for a fraction $f_\text{s}<0.6$ of Cu sites even at $T\sim0.01J$ is direct evidence for a spatially inhomogeneous local response in the presence of disorder in these materials, showing the close competition of various ground states.  These results also explain the seemingly contradictory interpretation of purely gapped or gapless behavior based on $^{17}$O NMR results~\cite{Fu2015,Khunita2020}, which did not properly account for inhomogeneity.

$1/T_{1}$ is a local probe of spin fluctuations at the resonant frequency $\omega_{\text{o}}$; $1/T_{\text{1}} \propto T|A_{\text{hf}}|^{2} \chi''(\omega_{\text{o}})/\omega_{\text{o}}$, where $A_{\text{hf}}$ is the hyperfine coupling between the observed nuclear spin and electron spin, and $\chi''(\omega_{\text{o}})$ is the imaginary part of the local spin susceptibility.  In general, $1/T_{1}\propto|A_{\text{hf}}|^{2}/J$ for Heisenberg antiferromagnets at high temperatures~\cite{Moriya1956II}, and $1/T_{1}$ remains roughly constant unless strong short range order enhances $1/T_{1}$ or a spin-gap suppresses $1/T_{1}$.  The spatial proximity between $^{63}$Cu nuclear spin and the spin-1/2 moment always enhances $1/T_{1}^\text{Cu}$ measured at paramagnetic Cu$^{2+}$ sites to large values $10^{3\sim4}$~s$^{-1}$ in Heisenberg antiferromagnets (e.g. \cite{Itoh1990,Azuma1994,Kikuchi1994,Kageyama1999}).  When Cu$^{2+}$ sites form magnetically inert singlet dimers favored by the non-frustrated geometry in two-leg spin-ladder SrCu$_2$O$_3$, spin-Peierls chain CuGeO$_3$, and Shastry-Sutherland lattice SrCu$_{2}$(BO$_{3}$)$_{2}$, $1/T_{1}^\text{Cu}$ is dramatically suppressed~\cite{Azuma1994,Kikuchi1994,Kageyama1999}.  

In analogy, if the kagome plane approaches a homogeneous QSL ground state, as schematically shown in Fig.~1{\bf a}, we expect {\it uniform suppression} of $1/T_{1}^{\text{Cu}}$, with or without a gap.  On the other hand, if spin singlets gradually emerge in the presence of disorder~\cite{Singh2010,Shimokawa2015,Kimchi2018,Kawamura2019},  as schematically shown in Fig.~1{\bf b}, $1/T_{1}^\text{Cu}$ would be suppressed only at {\it a temperature dependent fraction} $f_\text{s}$ of Cu sites.  Despite such clear-cut expectations, $1/T_{1}^{\text{Cu}}$~\cite{Imai2008} has been playing only a marginal role in the past debate, simply because the condensed matter NMR community did not know how to extract quantitative information about the large distributions of $1/T_{1}^{\text{Cu}}$. 

In order to test these hypotheses, we measured $1/T_{1}$ in zero external magnetic field at the peak of $^{63}$Cu and $^{79}$Br nuclear quadrupole resonance (NQR) lineshapes presented in Fig.~1{\bf d}, and $1/T_{1}$ at $^{19}$F sites in an external magnetic field of  0.72~T.  In Fig.~1{\bf g}, we summarize representative recovery curves $M(t)$.  We found that large distributions of $1/T_1$ make the single exponential fit of $M(t)$ unsatisfactory, especially below $30$~K.   To account for distributions of $1/T_1$, it is customary to fit $M(t)$ with a stretched exponential, $M(t) = M_\text{{o}}-A \cdot exp[-(ct/T_\text{{1,str}})^{\beta}]$ using a phenomenological stretched exponent $\beta$($<1$).  The pre-factor is $c=3$ for $^{63}$Cu and $^{79}$Br NQR (both with nuclear spin $I=3/2$), and $c=1$ for $^{19}$F NMR  ($I=1/2$).  See supplementary section I for additional examples of $M(t)$ and the temperature dependence of $\beta$.

In Fig.~2{\bf a-b}, we summarize the stretched fit $1/T_\text{{1,str}}^\text{{Cu}}$ measured with a short pulse separation time $\tau = 6$~$\mu$s ($+$ symbols).    $1/T_\text{{1,str}}^\text{{Cu}}$ is nearly identical between ZnCu$_{3}$(OD)$_{6}$FBr and ZnCu$_{3}$(OD)$_{6}$Cl$_2$ except below $\sim 4$~K, where fluctuating hyperfine magnetic fields originating from the larger concentration of Cu$^{2+}$ defect spins in ZnCu$_{3}$(OD)$_{6}$Cl$_2$ enhances the floor value of $1/T_\text{{1,str}}^\text{{Cu}}$.  The enhancement of $1/T_1$ observed for quadrupolar $^{63}$Cu and $^{79}$Br sites around 100~K in ZnCu$_{3}$(OD)$_{6}$FBr and around 50~K in ZnCu$_{3}$(OD)$_{6}$Cl$_{2}$ is caused by the slowing of fluctuating electric field gradient~\cite{Imai2008} induced by the structural distortion near the defects~\cite{Fu2015,Zorko2017}. 

Unlike the spin-singlet dimer materials~\cite{Azuma1994,Kikuchi1994,Kageyama1999}, $1/T_\text{1,str}^{\text{Cu}}$ hovers at fairly large values 10$^{2\sim3}$~s$^{-1}$ below $\sim$40~K, and hence does not hint at the presence of spin singlets.  But interpretation of  $1/T_{\text{1,str}}$  deduced from the conventional stretched fit requires ample precaution, because one can easily overlook even the presence of two distinct components of $1/T_1$~\cite{Singer2020}.  Without knowing the exact nature of the large distribution, it is risky to draw conclusions from $1/T_{\text{1,str}}$.  This is where NMR research into disordered quantum materials used to run into difficulties.

There is, however, a way to deduce the histogram $P(1/T_{1})$ of the distribution of $1/T_1$ directly from $M(t)$ by taking advantage of the ILTT$_1$ analysis technique based on Tikhonov regularization~\cite{Song2002,Singer2020,Papawassiliou2020}.  This technique was initially developed two decades ago for NMR research into petrophysics, but can be easily adopted to quantum materials~\cite{Singer2020,Papawassiliou2020}.  In the ILTT$_1$ analysis, one assumes only that $1/T_1$ is spatially distributed, and $M(t) = \sum_{j}{P(1/T_{1j})~\{1-2~exp(-ct/T_{1j})\}}$.  $1/T_{1j}$ is the j-th value of the distributed $1/T_1$, and $P(1/T_{1j})$ represents the corresponding {\it probability density} for the nuclear spins to relax with $1/T_{1j}$.  We numerically invert the discrete $M(t)$ data based on ILT, and deduce the histogram $P(1/T_{1})$ of the distributed $1/T_{1}$.  We refer readers to a brief review in section II of ref.\cite{Singer2020} and its Supplementary Materials for the established procedures of ILT, including how incomplete inversion of $M(t)$ is treated.

In Fig.~1{\bf g}, we compare the stretched and ILT fits of $M(t)$ for ZnCu$_{3}$(OD)$_{6}$FBr.  The poor stretched fit observed for $^{63}$Cu at 4.2~K due to the presence of a kink around $t\sim10^{-3}$~s is improved by the ILT fit.  In Fig.~3{\bf a}, we summarize the temperature evolution of $P(1/T_{1}^\text{{Cu}})$.  At 40~K, the distribution $P(1/T_{1}^\text{{Cu}})$ is centered around a single peak, marked as {\it para}, at $1/T_\text{{1,para}}^\text{{Cu}}\simeq 1000$~s$^{-1}$, a typical large value for the paramagnetic Cu$^{2+}$ sites.  Upon decreasing temperature, this main peak does not shift.  The constant $1/T_\text{{1,para}}^\text{{Cu}}$ suggests that some Cu sites remain paramagnetic with short spin-spin correlation length.  On the other hand, a distinct split-off peak, marked as {\it singlet}, gradually emerges with increasing weight.  In Fig.~2{\bf e}, we summarize the temperature evolution of $P(1/T_{1}^\text{{Cu}})$ in a color contour plot.  Notice that a light-blue ridge arising from the split-off peak emerges below about 30~K~($\sim 0.16J$), whereas the horizontal red section arising from the constant $1/T_\text{1,para}^\text{Cu}$ component progressively fades away. 

We deconvolute $P(1/T_{1}^\text{{Cu}})$ curves with two Gaussians to determine the location of the two peaks, $1/T_\text{{1,para}}^\text{{Cu}}$ and $1/T_\text{{1,singlet}}^\text{{Cu}}$.  We also determine the center of gravity $1/T_\text{1,cg}^\text{Cu}$ of the distribution of $P(1/T_{1}^\text{{Cu}})$, as marked with a bullet in Fig.~3{\bf a}.  In Fig.~2{\bf a}, we compare their temperature dependences with $1/T_\text{1,str}^\text{Cu}$.  Notice that $1/T_\text{1,str}^\text{Cu}$ is a reasonably good approximation of $1/T_\text{1,cg}^\text{Cu}$, but the analysis of $M(t)$ based on the former loses the information about the distribution of $1/T_{1}$.  In fact, the smoothly decreasing temperature dependence of $1/T_\text{1,str}^\text{Cu}$ does not reflect the individual behavior of $1/T_\text{{1,para}}^\text{{Cu}}$ and $1/T_\text{{1,singlet}}^\text{{Cu}}$.  Below $10$~K, $1/T_\text{1,singlet}^\text{Cu}$ is suppressed by nearly two orders of magnitude compared with $1/T_\text{1,para}^\text{Cu}$, and levels off at a floor value $\gtrsim 20$~s$^{-1}$.  This small value overestimates the intrinsic contributions from Cu$^{2+}$ spin fluctuations within the kagome plane, because the aforementioned Cu$^{2+}$ defect spins at Zn$^{2+}$ sites should induce  constant background contributions.   Such a small $1/T_{1}^\text{Cu}$ value of $\sim20$~s$^{-1}$ or less was previously observed for Cu$^{2+}$ sites only in the gapped spin-singlet ground state of spin dimer materials~\cite{Azuma1994,Kikuchi1994,Kageyama1999}.  Therefore, we can attribute the split-off peak of $P(1/T_{1}^\text{{Cu}})$ to the emergent singlets, whereas the para peak represents Cu$^{2+}$ sites that have not been involved in singlets yet.  ZnCu$_{3}$(OD)$_{6}$Cl$_{2}$ also exhibits remarkably similar results as shown in Figs.~2{\bf b}, 2{\bf f} and 3{\bf b}, except that the higher defect concentration in ZnCu$_{3}$(OD)$_{6}$Cl$_{2}$ results in smearing of the $P(1/T_{1}^\text{{Cu}})$ curves, and enhances the floor value of the singlet peak to $1/T_{\text{1,singlet}}^\text{{Cu}} \simeq 80$~s$^{-1}$.  
 
 The integral of the $P(1/T_{1})$ curves in Fig.~3 is normalized to 1, so that the total probability is 1.  Therefore, the integrated area under the purple dashed curve centered at $1/T_\text{1,singlet}^\text{Cu}$ represents the fraction $f_\text{s}$ of the Cu sites involved in the singlets.  We summarize the temperature dependence of $f_\text{s}$ in Fig.~4.  $f_\text{s}$ shows nearly identical behavior for ZnCu$_{3}$(OD)$_{6}$FBr and ZnCu$_{3}$(OD)$_{6}$Cl$_2$, and remains finite up to $\sim30$~K, suggesting that the break-up energy  $\Delta$ of singlets has a broad spatial distribution up to $\Delta \sim 30$~K.  A naive extrapolation of $f_\text{s}$ to $T=0$ may reach close to 100~\%.  But $f_{\text{s}}$ in Fig.~4 should be considered the upper bound, because we measured $1/T_\text{{1}}^\text{{Cu}}$ with a finite $\tau = 6$~$\mu$s, which tends to underestimate faster relaxing components.   We also note that $\mu$SR detected slow spin dynamics in these materials below $\sim6$~K~\cite{Tustain2020}, where some $^{63}$Cu and $^{79}$Br NQR signal intensity is lost, as shown in supplementary section I.  These findings suggest incipient freezing of some Cu$^{2+}$ spins.  Therefore, the $f_{\text{s}}$ results below $\sim6$~K represent the fraction of the singlet Cu$^{2+}$ sites in a region that are not about to freeze.  
 
The fact that $f_\text{s}$ does not exceed $60$\% even at $T\sim0.01J$ indicates that theoretical models derived for the pristine kagome Heisenberg antiferromagnet are insufficient to account for the low-energy properties of these kagome materials with disorder.  Instead, our $P(1/T_{1}^\text{Cu})$ results indicate that spin singlets gradually emerge with spatially varying gaps. A variety of interesting theoretical scenarios may result from the disorder~\cite{Kimchi2018,Singh2010,Shimokawa2015,Kawamura2019}, however, the presence of a spin-gap in part of the sample is clear.  While it is not straightforward to extrapolate this finding to the disorder free limit, it seems unlikely that the gap is induced by disorder, in view of the fact that disorder generally generates low lying excitations.

It is also worth noting that if these materials are a realization of a QSL with a uniform gap $\Delta$, $P(1/T_{1}^\text{Cu})$ should have {\it a single peak} representing 100~\% of Cu sites, and the peak $1/T_{1}^\text{Cu}$ value should exhibit an activated behavior for that uniform gap $\Delta$.  We also note that the apparent linear behavior of $1/T_\text{1,str}^\text{Cu}$ in the log-log plot Fig.~2{\bf b} could be misinterpreted as a power-law behavior of $1/T_\text{1}^\text{Cu}\sim T^{\alpha}$ ($\alpha<1$) expected for Dirac fermions in gapless spin liquids~\cite{Ran2007}, but it is illusory and disappears when confronted with the actual distribution of $1/T_{1}^\text{Cu}$. In reality, neither $1/T_\text{1,para}^\text{Cu}$ nor $1/T_\text{1,singlet}^\text{Cu}$ obeys such a power law for both ZnCu$_{3}$(OD)$_{6}$Cl$_2$ and ZnCu$_{3}$(OD)$_{6}$FBr.  Therefore, the fully gapless QSL scenario advocated earlier based on an apparent power law behavior of $1/T_{1}^\text{O}$ observed for $^{17}$O sites~\cite{Khunita2020} is ruled out by $P(1/T_{1}^\text{Cu})$. 

NMR data at other sites provide additional clues about the nature of the inhomogeneously gapped regions of the kagome planes.  As shown in Fig.~1{\bf f}, $^{79}$Br and $^{19}$F sites are equidistant with 6 and 12 Cu$^{2+}$ sites, respectively, and hence $1/T_{1}^\text{Br,F}$ probe their average behavior. If the spin singlets form a valence bond solid consisting of large clusters as depicted in Fig.~1{\bf c}, some $^{79}$Br and/or $^{19}$F sites would be surrounded by magnetically inert spins and exhibit a split-off peak in $P(1/T_{1}^\text{Br,F})$.  The absence of such a signature in $P(1/T_{1}^\text{{Br,F}})$ suggests that  $^{79}$Br and $^{19}$F sites are surrounded randomly by both spin singlets and paramagnetic spins, as opposed to the dimer order expected for a valence bond crystal.  Earlier $^{17}$O NMR results~\cite{Fu2015,Khunita2020} also corroborate the present findings.  Notice that each $^{17}$O site is sandwiched between two Cu spins; when a certain fraction $f_\text{s}$ of Cu spins form the singlets below $\sim30$~K, a corresponding fraction of $^{17}$O sites are within the singlets and become magnetically inert.  It naturally explains why $^{17}K$ defined at the center peak of the extremely broad $^{17}$O NMR line begins to exhibit a gapped behavior below $\sim30$~K~\cite{Fu2015,Khunita2020} with a small gap $\Delta\simeq10$~K~\cite{Fu2015}.  This is also supported by the gapped behavior of $1/T_\text{1}^\text{O}$ observed for the second upper satellite quadrupole transition of $^{17}$O sites, which was narrowed and isolated by using a low magnetic field of 3.2~T to avoid contamination by other quadrupole satellites~\cite{Fu2015}.  The observed gap at $^{63}$Cu and $^{17}$O sites is attributed to the break-up energy of the singlets and other local excitations, and does not represent a uniform gap throughout the kagome planes.   In fact, a fraction $1-f_\text{s}$ of Cu sites remain paramagnetic with constant $1/T_\text{1,para}^\text{Cu}$ down to $T\sim 0.01J$, and these Cu sites lack a spin excitation gap.  This is also in agreement with the recent $^{17}K$ results determined for $^{17}$O sites with very fast relaxation rates~\cite{Khunita2020}.  Further microscopic theories based on the specific disorder present in the materials are necessary to fully understand the quantitative aspect of $f_{s}$, and the origins of the observed gap and its spatial distribution.\\


\begin{center}
{\bf Acknowledgements}\\
\end{center}
T.I. thanks T.~Sakai, K.~Uematsu, R.~Singh, I.~Kimchi, P.~A.~Lee, and S.~Sachdev for helpful communications, and P.~Dube and R.~Giannetta for technical assistance.  The work at McMaster was supported by NSERC (T.I.).  P.M.S. was supported by the Rice University Consortium for Processes in Porous Media.  The work at Stanford and SLAC (sample synthesis and characterization) was supported by the U.S. Department of Energy (DOE), Office of Science, Basic Energy Sciences, Materials Sciences and Engineering Division, under contract no. DE-AC02-76SF00515 (Y.S.L.).  R.W.S. was supported by a NSF Graduate Research Fellowship (DGE-1656518). \\

\begin{center}
{\bf Author contributions}\\
\end{center}
T.I. and Y.S.L. conceived the project.  R.W.S., W.H., J.~Wen and Y.S.L. synthesized and characterized the samples.  J.~Wang, W.Y., P.M.S. and T.I. carried out the NMR measurements and data analysis.  All authors contributed to the writing and editing of the manuscript.\\

\begin{center}
{\bf Competing interests}\\
\end{center}
The authors declare that they have no competing interests.\\

\begin{center}
{\bf Supplementary information}\\
\end{center}
Supplementary Sections I-IV, and Figs.~S1 - S9.\\

\begin{center}
{\bf METHODS}\\
\end{center}
\indent We synthesized the deuterated (D = $^{2}$H) powder samples of ZnCu$_{3}$(OD)$_{6}$FBr  and  ZnCu$_{3}$(OD)$_{6}$Cl$_2$  based on the procedures described in detail in \cite{Smaha2020} and \cite{Shores2005}.  We confirmed the sample quality based on powder X-ray diffraction measurements.  We also confirmed the interlayer defect concentrations and absence of observable in-plane defects based on site-selective X-ray studies and ICP chemical analysis for samples with the same synthesis batches used in this study.~\cite{Smaha2020_PRM}.  We conducted the spin echo NMR and NQR measurements using standard pulsed NMR spectrometers.  The typical pulse width is $\sim2.5$ and $\sim5$~$\mu$s for the 90$^{\circ}$ and 180$^{\circ}$ radio frequency pulses, respectively.  We measured the NQR lineshapes in Fig.~1{\bf d} using a fixed pulse separation time $\tau=15$~$\mu$s, and normalized the spin echo intensity for the sensitivity by dividing with the frequency squared.  We assigned the NQR peaks of $^{63}$Cu $^{65}$Cu, $^{79}$Br and $^{81}$Br isotopes based on the ratio of the peak frequencies, which is proportional to the ratio of the nuclear quadrupole moment, $^{63}Q/^{65}Q$ = 0.927 and $^{81}Q/^{79}Q$ = 0.837.  We conducted all the $1/T_1$ measurements by applying an inversion pulse prior to the spin echo sequence.  For $1/T_{1}^{\text{Cu}}$ measurements, we used a short pulse separation time $\tau = 6$~$\mu$s between the 90 and 180 degree pulses unless otherwise noted.  We conducted the ILT analysis of $M(t)$ based on the established procedures described in \cite{Singer2020}.  We also confirmed the validity of the ILT results through two additional checks: (i) the two component analysis of $M(t)$ in the time domain leads to consistent results, and (ii) $1/T_{1}^\text{{Cu}}$ measured with a long $\tau=30$~$\mu$s agrees with $1/T_\text{{1,singlet}}^\text{{Cu}}$.  See supplementary section III and IV for additional details.  We chose to conduct $^{19}$F NMR measurements of $1/T_{1}^{\text{F}}$ in an external magnetic field of 0.72~T, so that the $^{19}$F resonant frequency coincides with the $^{79}$Br NQR peak frequency in zero field.  This is because the enhancement of $1/T_{1}$ due to fluctuating electric field gradient at $^{35}$Cl sites of ZnCu$_{3}$(OH)$_{6}$Cl$_2$ is known to depend mildly on frequency~\cite{Imai2008}.  The absence of a peak in $1/T_{1}^{\text{F}}$ around 100~K in Fig.~2{\bf d} for non-quadrupolar $^{19}$F sites, as well as the isotope ratio of $1/T_{1}^{\text{Br}}$ between $^{79}$Br and $^{81}$Br in supplemental section II, confirms that the large enhancement of $1/T_\text{{1,str}}^\text{{Br}}$ around 100~K is quadrupolar in nature, and extrinsic to spin physics.  We confirmed that $^{19}$F Knight shift measured  at 0.72~T agrees well with the higher field results above $\sim10$~K~\cite{Feng2017}.  But the dramatic line broadening in 0.72~T overwhelms the small Knight shift at lower temperatures and makes accurate determination of the latter difficult. \\

\begin{center}
{\bf Data availability}\\
\end{center}
The data sets generated during the current study are available from the corresponding author on reasonable request.\\


\begin{figure}
\centering
\includegraphics[width=6in]{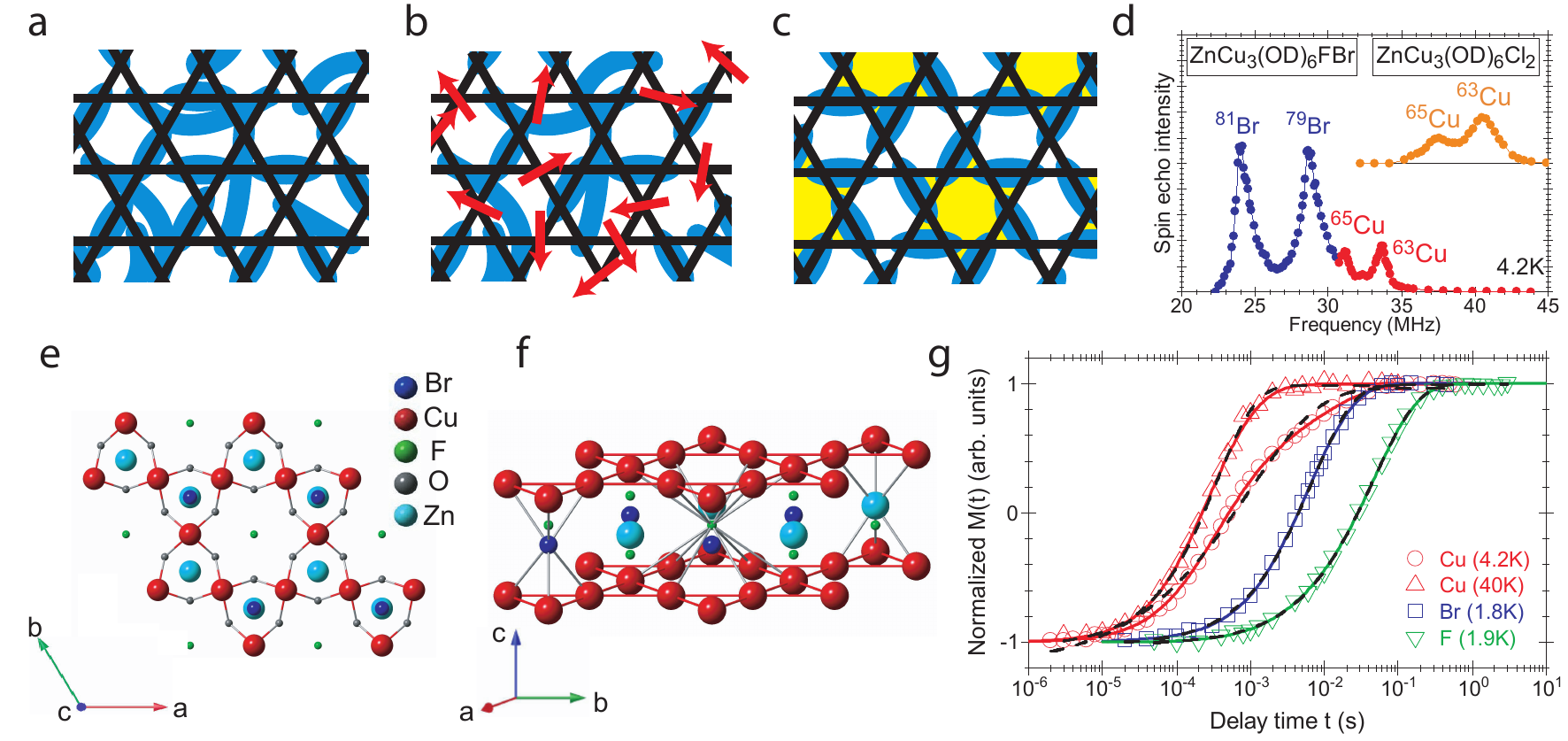}
\caption{{\bf Fig.1.  Spin singlets in the kagome lattice}: {\bf a} A conceptual snap shot of the fluctuating spin singlets (blue ovals) on the pristine kagome lattice forming an ideal quantum spin liquid state.  {\bf b} A mixture of spin singlets and paramagnetic spins (red arrows), realized in the inhomogeneous kagome planes of our materials with disorder.  {\bf c} An example of the valence bond solid consisting of a fixed pattern of 6 spin singlets forming a pinwheel structure (yellow).  {\bf d} $^{63,65}$Cu and  $^{79,81}$Br NQR lineshapes in ZnCu$_{3}$(OD)$_{6}$FBr.  Also shown with a vertical offset is the $^{63,65}$Cu NQR lineshapes in ZnCu$_{3}$(OD)$_{6}$Cl$_{2}$.  {\bf e} c-axis view (D sites are not shown) and {\bf f} a tilted view (D and O sites are not shown) of the crystal structure of ZnCu$_{3}$(OD)$_{6}$FBr .    {\bf g} Examples of the normalized $M(t)$ observed for ZnCu$_{3}$(OD)$_{6}$FBr, with stretched (black dashed line) and ILT (solid line) fits.
}
\label{crystal}
\end{figure}

\begin{figure}
\centering
\includegraphics[width=6in]{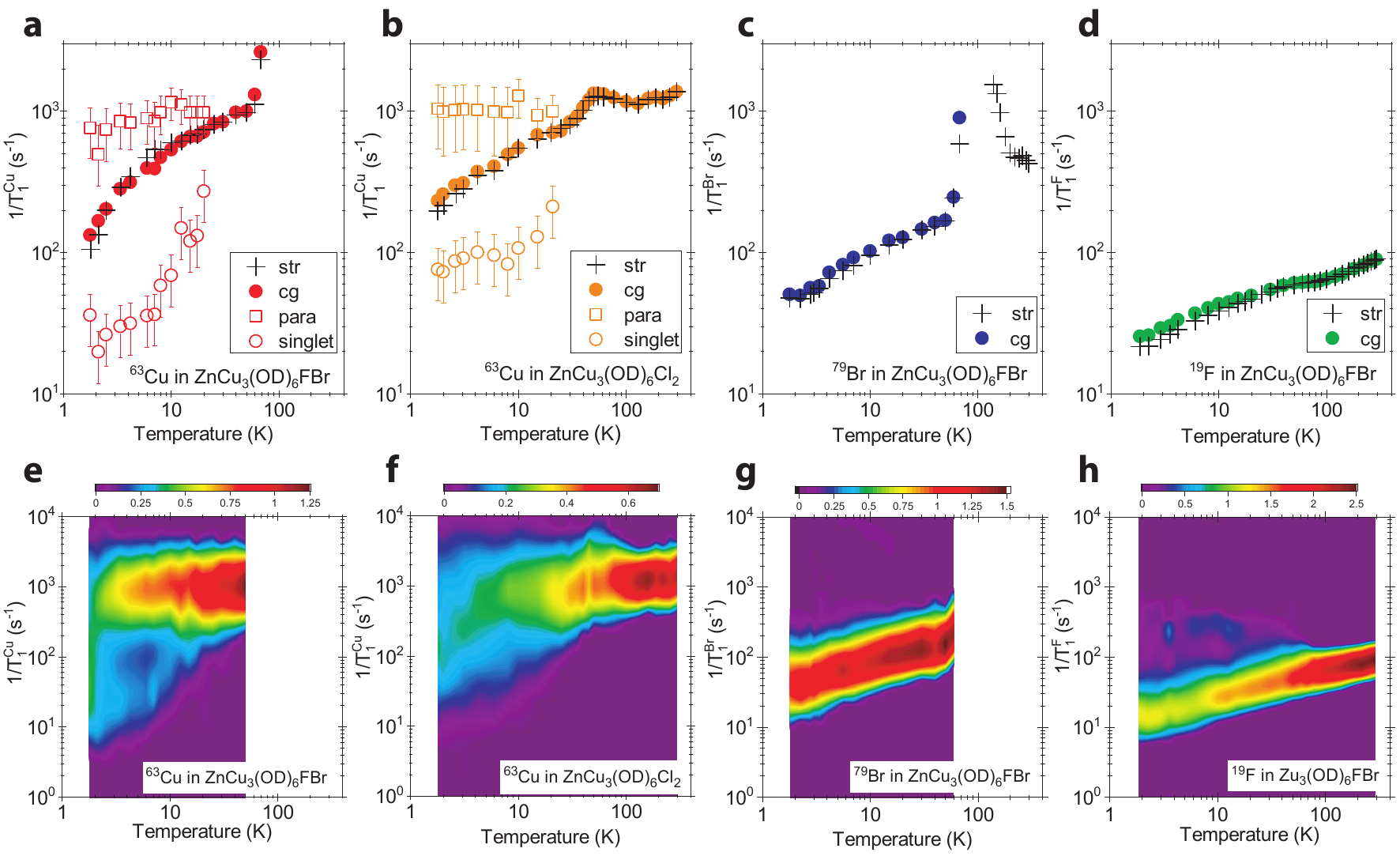}
\caption{{\bf Fig.2.  $1/T_{1}$ and its distribution $P(1/T_{1})$}: {\bf a-d} (+) $1/T_\text{1,str}$ as determined from the conventional stretched fit of $M(t)$ at {\bf a} $^{63}$Cu, {\bf c} $^{79}$Br, and {\bf d} $^{19}$F sites in ZnCu$_{3}$(OD)$_{6}$FBr, and {\bf b} $^{63}$Cu sites in ZnCu$_{3}$(OD)$_{6}$Cl$_{2}$.  Also compared in {\bf a-d} are: ($\bullet$) center of gravity $1/T_\text{1,cg}$, ($\square$) paramagnetic $1/T_\text{1,para}^\text{Cu}$, and ($\circ$) singlet $1/T_\text{{1,singlet}}^\text{{Cu}}$ as determined from the density distribution $P(1/T_{1})$.  The error bars for $1/T_\text{1,para}^\text{Cu}$ and $1/T_\text{{1,singlet}}^\text{{Cu}}$ represent the absolute maximum/minimum in the peak locations of the double Gaussian fit of $P(1/T_{1})$.  {\bf e - h} The color contour plot of $P(1/T_{1})$ corresponding to the results presented in upper panels {\bf a - d}.  The color scales are shown in the horizontal bar above each panel.
}
\label{crystal}
\end{figure}

\begin{figure}
\centering
\includegraphics[width=6in]{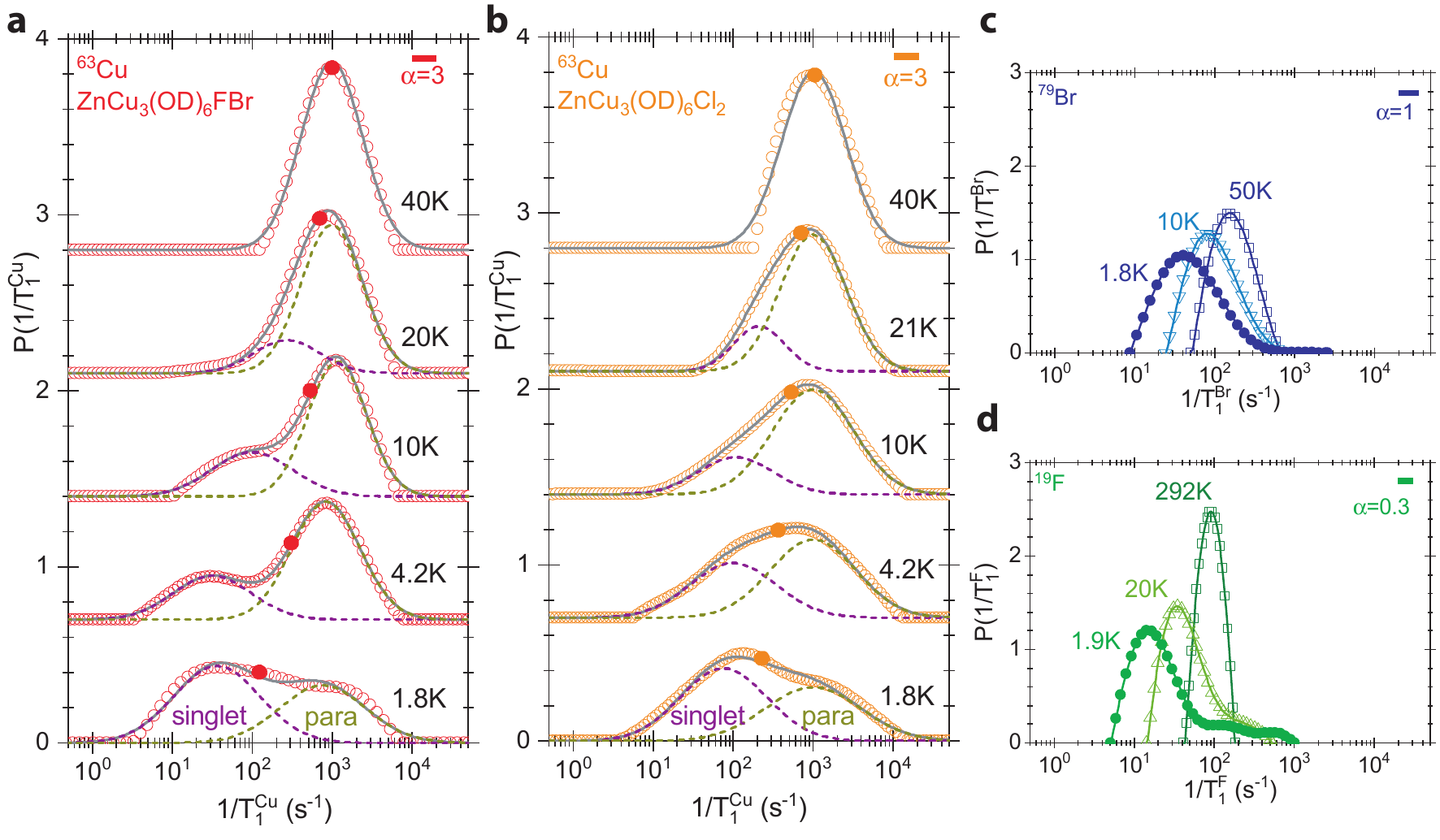}
\caption{
{\bf Fig.3.  Distribution of the Cu spin environment}: $P(1/T_{1})$ deduced by ILT for {\bf a} $^{63}$Cu, {\bf c} $^{79}$Br, and {\bf d} $^{19}$F sites in ZnCu$_{3}$(OD)$_{6}$FBr, and {\bf b} $^{63}$Cu sites in ZnCu$_{3}$(OD)$_{6}$Cl$_{2}$. $\bullet$ overlaid on $P(1/T_{1}^\text{{Cu}})$ in {\bf a} and {\bf b} mark the center of gravity of $P(1/T_{1})$ on a log scale, gray curves represent the single (at 40~K) or double Gaussian fit, and the dashed curves are the deconvolution to the singlet and paramagnetic (para) contributions based on the double Gaussian fit. For clarity, the vertical origin in {\bf a} and {\bf b} is offset at 4.2~K and above.  The thick horizontal bar at the upper right corner of each panel is the typical additional width introduced by the ILT itself for the fixed value of the regularization parameter $\alpha$ (i.e. a crude measure of the effective resolution of $P(1/T_{1})$; see supplementary section IV for details).
}
\label{crystal}
\end{figure}

\begin{figure}
\centering
\includegraphics[width=4in]{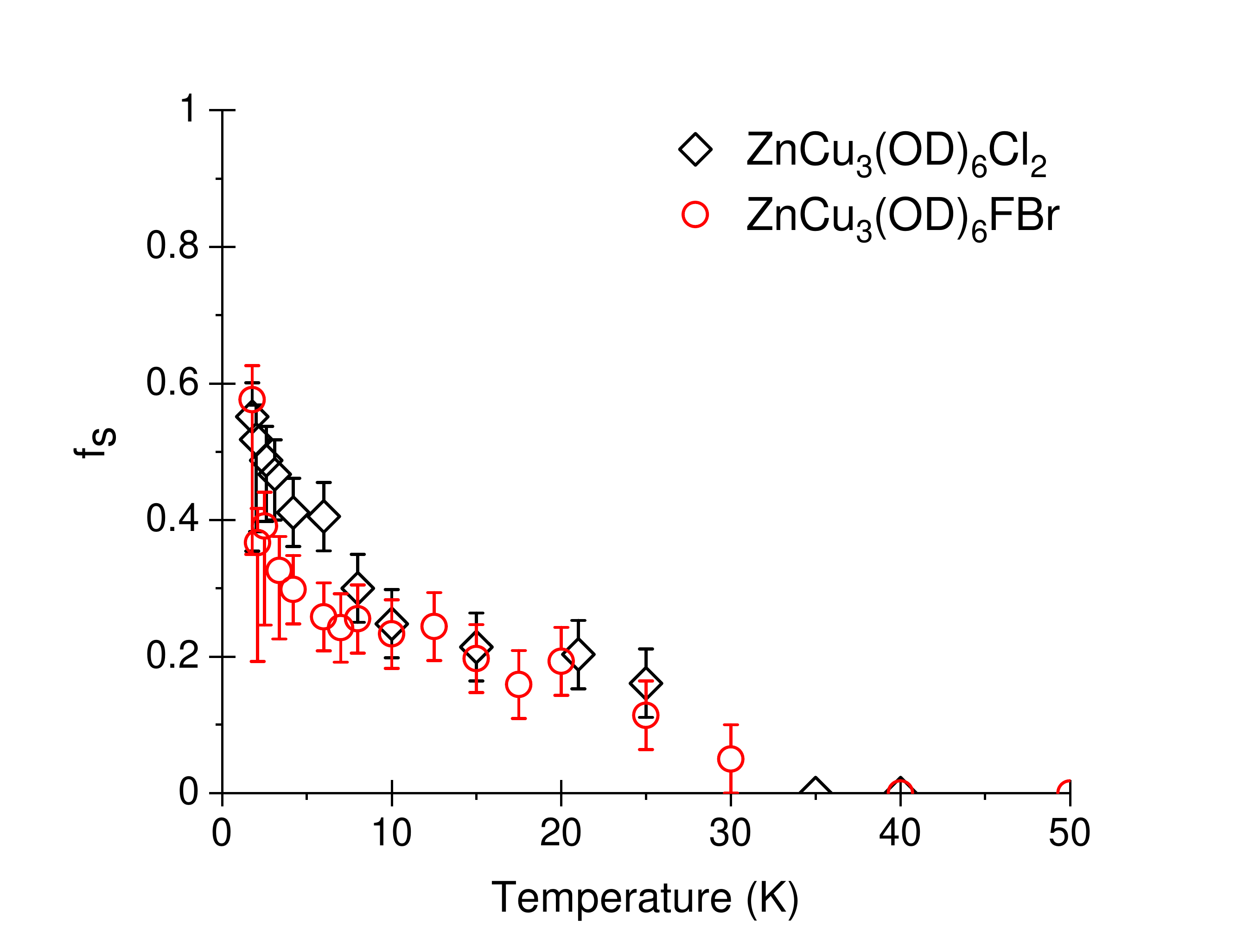}
\caption{
{\bf Fig.4.  Singlet fraction}: The fraction of Cu site involved in spin singlets, $f_\text{s}$, estimated for $\tau = 6$~$\mu$s from the integral of the purple dashed curves in Fig.3{\bf a} and {\bf b} for ZnCu$_{3}$(OD)$_{6}$FBr (red $\circ$) and ZnCu$_{3}$(OD)$_{6}$Cl$_{2}$ (black $\diamond$).  The lower error bars below 6~K represent the lower bound of $f_\text{s}$, estimated based on the assumption that the $^{63}$Cu signal loss affects only the paramagnetic component.  All other error bars correspond to the absolute maxima or minima.
}
\label{crystal}
\end{figure}

\newpage


%

\end{document}